# Algorithmic Inheritance: Surname Bias in AI Decisions Reinforces Intergenerational Inequality

Pat Pataranutaporn

*Massachusetts Institute of Technology, Cambridge, MA, USA*

Nattavudh Powdthavee

*Nanyang Technological University, Singapore, and IZA, University of Bonn, Germany*

Pattie Maes

*Massachusetts Institute of Technology, Cambridge, MA, USA*

# Abstract

Surnames often convey implicit markers of social status, wealth, and lineage, shaping perceptions in ways that can perpetuate systemic biases and intergenerational inequality. This study is the first of its kind to investigate whether and how surnames influence AI-driven decision-making, focusing on their effects across key areas such as hiring recommendations, leadership appointments, and loan approvals. Using 72,000 evaluations of 600 surnames from the United States and Thailand—two countries with distinct sociohistorical contexts and surname conventions—we classify names into four categories: "Rich," "Legacy," "Normal," and phonetically similar "Variant" groups. Our findings show that elite surnames consistently increase AI-generated perceptions of power, intelligence, and wealth, which in turn influence AI-driven decisions in high-stakes contexts. Mediation analysis reveals perceived intelligence as a key mechanism through which surname biases influence AI decision-making process. While providing objective qualifications alongside surnames mitigates most of these biases, it does not eliminate them entirely—especially in contexts where candidate credentials are low. These findings highlight the need for fairness-aware algorithms and robust policy measures to prevent AI systems from reinforcing systemic inequalities tied to surnames—an often-overlooked bias compared to more salient characteristics such as race and gender. Our work calls for a critical reassessment of algorithmic accountability and its broader societal impact, particularly in systems designed to uphold meritocratic principles while counteracting the perpetuation of intergenerational privilege.

**Introduction**

Persistent inequalities in labor markets are well documented: employers frequently engage in discriminatory practices based on gender, race, and immigration status—whether intentionally or inadvertently (1–3). With the rise of AI-driven recruitment, these biases are assuming new forms. Algorithms trained on historical data risk replicating or even amplifying existing inequalities, as exemplified by Amazon's discontinued hiring tool that systematically disadvantaged women (4–6). Such evidence has intensified scrutiny over the subtle cues—both explicit and implicit—that may trigger biased responses in hiring and other high-stakes decisions. Although attention to race and gender bias in algorithmic systems is increasing (7–9), little research has focused on surname bias. Surnames can serve as covert proxies for inherited socioeconomic status, enabling class-based discrimination to persist even in AI systems designed to be race- and gender-neutral. This suggests that eliminating explicit demographic markers does not necessarily eliminate bias, as surnames may continue to encode and perpetuate historical inequalities. Given these concerns, a systematic investigation into surname bias is critical for understanding and mitigating the multifaceted dimensions of discrimination in AI-driven decision making.

A substantial body of research has examined discrimination based on first names, as demonstrated by numerous field experiments on labor market bias (2, 10–12). In contrast, bias linked to surnames has received relatively little attention despite its unique and far-reaching implications. Unlike first names—which individuals can often change relatively easily—surnames are typically inherited or acquired through marriage. As enduring markers of lineage, surnames encapsulate signals of wealth, privilege, and social status, reflecting historical advantages that persist across generations (13–15). Economic historian Gregory Clark, for example, has shown that tracking surnames over time reveals a disproportionate representation of certain elite surnames among the wealthiest and most influential groups, even amid significant social and economic change (13, 14). Clark argues that surnames not only serve as proxies for inherited advantages but also signal a family's legacy of cultural and social capital—non-economic assets such as education, skills, knowledge, values, and even tastes and preferences—that families acquire and transmit, along with social capital—the networks of relationships and connections built over time—and cumulative advantages derived from innate abilities, acquired behaviors, or even genetic predispositions, all of which are passed down through generations..

These historical insights have clear implications for contemporary AI-driven decision-making. Like education in signaling models (16), where people use someone's qualification as a signal for their unobserved abilities and intelligence in the job market, elite surnames signal a privileged family background and access to greater resources. As a result, individuals bearing such surnames may be perceived as highly capable and intelligent, receiving preferential treatment in professional and social settings. This signaling effect reinforces cycles of advantage, as societal recognition and validation of elite surnames further entrench intergenerational benefits. Yet, despite these significant implications, the extent to which status-laden surnames influence outcomes in AI-driven decision-making processes remains poorly understood—especially given mounting concerns that algorithmic models may encode and amplify historical biases.

This study investigates whether and how surnames—often presumed impartial—trigger discrimination in AI-based decision-making. Our comprehensive approach yielded 72,000 evaluations generated from 600 surnames assessed across 10 dimensions with 3 evaluations per dimension and 4 candidate profile configurations (good, medium, bad, and retracted). We measured the effects of these surnames on outcomes in key domains, including hiring recommendations, leadership appointments, and loan approvals, using GPT-4o-mini, the leading language model in 2024. Data collection was performed via a standardized API-based protocol to ensure rigor and reproducibility.

To broaden the analytical scope, we drew on surnames from two distinct national contexts. To ensure a broad and comparative perspective, we draw on surnames from the United States (N=300) and Thailand (N=300)—two countries with significant socioeconomic diversity, distinct historical contexts of intergenerational inequality, and documented institutional discrimination linked to family names (17–22). We categorize surnames into four groups. "Rich Surnames" are drawn from the Forbes lists of wealthiest individuals in 2024 and 2025 in each country. "Legacy Surnames" include royally bestowed names in Thailand and prominent dynastic family names in the U.S. "Normal Surnames" are randomly sampled from each country's most commonly occurring names. Finally, "Variant Surnames" resemble both elite and normal surnames phonetically but differ in spelling, allowing us to test whether similarity alone triggers bias in AI models.

Building on this comparative framework, we also contextualize our findings within unique legal and demographic settings. Focusing on the U.S. provides a critical lens for AI development and deployment, as American-trained models often see global application. Thailand, on the other hand, offers the distinctive context of its 1913 Surname Act, which mandates unique surnames for every family. This legal specificity allows us to isolate how lineage-based attributes shape algorithmic decisions without the confounding overlap commonly encountered in countries like the U.S., where multiple families share the same surname. Furthermore, Thailand's relatively racially homogeneous population—with the majority of citizens being of Southeast or East Asian descent—contrasts sharply with the more racially diverse U.S., enabling a nuanced analysis of surname effects across different societal contexts. By analyzing how elite versus ordinary surnames influence AI assessments in these contrasting settings, our study provides timely insight into the extent to which algorithmic decision-making might perpetuate or mitigate inherited inequalities.

Finally, our investigation contributes to broader efforts to promote fairness in AI systems. By examining the algorithmic processes behind high-stakes decisions, our work not only uncovers the hidden persistence of intergenerational inequality linked to surnames but also provides a foundation for designing more equitable AI systems. In doing so, the study reveals how immutable surname markers—deeply intertwined with historical legacies of privilege—can subtly bias outcomes in employment, leadership, and financial services. These insights are crucial for policymakers, AI developers, and stakeholders striving to design and implement systems that are both fair and transparent. By advancing the academic discourse on algorithmic fairness and contributing to broader societal efforts to dismantle enduring structures of inequality, our research underscores the long-term consequences of past discrimination in the realm of modern technology.

**Results**

To assess potential biases associated with surnames, we conducted systematic evaluations using GPT-4o-mini, the industry-leading language model in 2024. Each surname was evaluated across ten dimensions relevant to socioeconomic perception and decision-making, combined with four candidate profile configurations, resulting in a total of 72,000 evaluations. Scores were assigned on a scale from 0 to 10.

We then employed ordinary least squares (OLS) regression to estimate three distinct models examining the influence of surname categories on AI assessments and the resulting recommendations. First, we analyzed how the AI evaluates surnames—across dimensions such as power, wealth, intelligence, and commonality—based on categories including legacy, legacy variants, rich, rich variants, and common variants (see Figure 1). Next, we linked these surname categories directly to AI-generated recommendations for outcomes such as executive hiring, leadership appointments, entry-level hiring, international school admissions, political careers, and loan approvals (Figure 2). Finally, we integrated these approaches by regressing recommendations on both AI perceptions and surname categories to reveal the mediating role of perceptions in driving outcomes (Figures 3 and 4).

To ensure the robustness of our findings, we applied bootstrap standard errors with 1,000 replications, conducted Sobel-Goodman mediation tests via structural equation modeling using STATA's *medsem* command, and adjusted p-values for multiple comparisons with STATA's *wyoung* command.

Our study shows that in both the U.S. and Thailand, legacy and rich surnames significantly shape AI assessments of wealth and intelligence—effects that, in turn, influence consequential real-world decisions. Specifically, these elite surnames are linked to significant increases in perceived wealth and intelligence. The impact on perceived power, however, is country-dependent: while significant effects are observed in Thailand, they are absent in the U.S. Moreover, the enhanced perceptions associated with elite surnames partially mediate favorable AI recommendations for executive hiring, leadership appointments, and educational opportunities, primarily through perceptions of intelligence and, in some contexts, power. Although providing objective qualifications alongside surnames attenuates these biases, it does not fully eliminate them, particularly when candidate credentials are low.

**Surname Bias on AI Perceptions of Wealth, Intelligence, Power, and Commonality**

First, we examined how the AI evaluates surnames along dimensions of power, wealth, intelligence, and commonality across various categories—including legacy, legacy variants, rich, rich variants, and common variants. Figure 1 summarizes our primary findings by presenting box plots and ordinary least squares coefficient plots that detail AI evaluations of perceived power, wealth, intelligence, and commonality for each surname category.

Consistent with Clark's work on the link between surnames and intergenerational inequality (13, 14), we find that surnames also signal wealth and status in AI judgments. In our study, in both the U.S. and Thailand, legacy and rich surname categories emerge as robust predictors of perceived wealth. In Thailand, legacy surnames are associated with a 0.674-point increase in perceived wealth ($p < 0.001$, 95% CI [0.325, 1.022]) relative to common surnames, while rich surnames correspond to a 0.893-point increase ($p < 0.001$, 95% CI [0.549, 1.236]). This relationship is even more pronounced in the U.S., where legacy surnames yield a 2.246-point increase ($p < 0.001$, 95% CI [1.693, 2.800]) and rich surnames a 1.926-point increase ($p < 0.001$, 95% CI [1.328, 2.525]). These correlations are considerable, particularly when contextualized by the mean perceived power levels, which are 5.38 (SD = 0.84) in Thailand and 2.95 (SD = 0.85) in the U.S.

Also consistent with signaling models in the labor market (15), legacy and rich surnames significantly predict perceived intelligence in both countries. In Thailand, legacy surnames are associated with a 0.660-point increase in perceived intelligence ($p < 0.001$, 95% CI [0.461, 0.860]), while rich surnames correspond to a 0.620-point increase ($p < 0.001$, 95% CI [0.409, 0.831]). In the U.S., legacy surnames yield a 0.386-point increase in perceived intelligence ($p = 0.038$, 95% CI [0.021, 0.751]), whereas rich surnames exhibit a larger increase of 0.632 points ($p < 0.001$, 95% CI [0.295, 0.970]).

Legacy and rich surnames are positively and statistically significantly associated with perceived power in Thailand but not in the U.S. However, these associations are weaker compared to those observed for perceived wealth and intelligence, thus suggesting that while elite surnames may carry an implicit status signal, their influence on perceptions of power varies by cultural context.

Moreover, in both countries, AI consistently rates all elite surnames and their variants as significantly less common than those in the "Common surnames" category. For example, legacy surnames are associated with a decrease of 1.581 points ($p < 0.001$, 95% CI [-1.936, -1.225]) in perceived commonality in Thailand and 3.620 points (p < 0.001, 95% CI [-4.250, -2.990]) in the U.S. On the other hand, rich surnames are associated with a decrease of 1.720 points (p < 0.001, 95% CI [-2.115, -1.324]) in perceived commonality in Thailand and 3.953 (p < 0.001, 95% CI [-4.587, -3.319]) in the U.S. This reinforces their distinctive and prestigious nature, further highlighting how rarity itself may contribute to the perception of exclusivity and status.

**Surname Bias Influences AI Decisions with Real-World Outcomes**

We investigate further whether legacy and rich surnames, along with their variances, affect AI judgments and decisions regarding outcomes with potential real-world consequences. The coefficient plots presented in Figure 2—and the raw data distribution illustrated in Figures 1A and 2A in the SI—take the investigation a step further. Looking across columns, we observe that legacy and rich surnames exert a positive and statistically significant influence on several, though not all, AI judgments. For instance, in Thailand, legacy surnames are associated with a 0.140-point increase in AI recommendations for executive hire (p = 0.031, 95% CI [0.012, 0.267]). In the U.S., rich surnames are associated with a 0.124-point increase in AI recommendations for executive hire (p < 0.001, 95% CI [0.200, 0.680]) and a 0.374-point increase in leadership rating (p < 0.001, 95% CI [0.165, 0.582]). Legacy surnames are also associated with a 0.221-point increase in leadership rating (p < 0.001, 95% CI [0.014, 0.427]). By contrast, legacy and rich surnames are both negatively associated with AI recommendations for entry hire in the U.S. The coefficient for legacy surnames in the entry hire regression is -0.567 (p < 0.001, 95% CI [-0.884, -0.251]), while the coefficient for rich surnames is -0.507 (p < 0.001, 95% CI [-0.760, -0.253]).

**The influences of real versus variant surnames**

Interestingly, AI also evaluated legacy and rich variants—surnames that phonetically resemble legacy and rich surnames but differ in spelling—positively in terms of perceived intelligence. In Thailand, for example, legacy variants are associated with a 0.507-point increase in perceived intelligence (p < 0.001, 95% CI [0.311, 0.703]) compared to common surnames (see Figure 1). These correlations are substantial, given that the mean perceived intelligence scores are 7.08 (S.D. = 0.68) in Thailand and 6.28 (S.D. = 1.11) in the U.S. This suggests that even minor variations in elite surnames can serve as implicit status signals, significantly influencing AI-driven assessments of intelligence.

However, the evidence supporting the positive influence of variant surnames on real-world outcomes is weaker. In the U.S., legacy and rich variants are positively associated only with recommendations for a political career. In the political career regression, the coefficient for legacy variants is 0.401 (p = 0.006, 95% CI [0.116, 0.685]), while the coefficient for rich variants is 0.573 (p < 0.001, 95% CI [0.297, 0.849]). Instead, legacy and rich variants are more often negatively and statistically significantly associated with real-world outcomes, including executive hiring, entry hiring, and international school admission in the U.S. Similarly, in Thailand, the coefficients for rich variants were negative and statistically significant in leadership, entry hiring, international school admission, and political career regressions. This suggests that while these variants may shape AI's perception of intelligence, their positive effect does not extend to real-world outcomes in the same way as genuine legacy and rich surnames.

**Mediation analysis identifies perceived intelligence as a crucial pathway through which surname biases operate**

The results from Figures 1 and 2 indicate that not all legacy and rich surnames strongly predict AI judgments in real-world scenarios. However, certain surnames that shape AI perceptions of power, wealth, intelligence, and commonality may still exert a meaningful influence.

To investigate this further, Figure 3 presents the partial correlations between AI perceptions of surnames and AI recommendations across various domains, controlling for surname categories within the same regression model. The findings reveal that perceived intelligence is a strong predictor of AI-driven recommendations for executive hiring, entry-level hiring, leadership roles, international school admissions, and loan approvals in both Thailand and the U.S. These results remain robust even after controlling for surname category fixed effects within the regression model.

For instance, a one-unit increase in perceived intelligence is associated with a 0.160-point increase (p < 0.001, 95% CI [0.071, 0.248]) in AI recommendations for executive hiring in Thailand, and a 0.527-point increase (p < 0.001, 95% CI [0.364, 0.690]) in the U.S. Similarly, perceived intelligence based solely on surnames predicts AI recommendations for international school admissions, a marker of educational elitism—particularly in the Thai context—demonstrating how AI assessments of surname-driven perceptions can influence high-stakes decision-making in both countries.

Perceived power strongly predicts AI recommendations for entry-level hiring in the U.S., leadership roles in Thailand, political careers in both Thailand and the U.S., and loan approvals in the U.S. Compared to perceived intelligence and power, perceived wealth has minimal predictive influence on these six outcomes. Notably, perceived power is strongly associated with a reduction in AI recommendations for entry-level hiring (=-0.172, p < 0.001, 95% CI [-0.265, -0.079]), suggesting that the wealthier AI perceives the surname to be, the less likely it is to recommend the individual for an entry-level position. Finally, perceived commonality of surnames only strongly predicts AI recommendation of a political career in Thailand. These findings support our hypothesis that not all legacy and wealthy surnames influence AI recommendations. Only those surnames that significantly impact AI ratings of perceived power and, particularly, intelligence, exhibit such effects.

For completeness, Figure 4 presents the partial correlations between surname categories and AI-generated recommendations across real-world outcomes. Legacy and rich surnames now demonstrate even more limited predictive power for most of these six outcomes, indicating that surname influence on AI judgments largely stems from perceptions of power and intelligence associated with the surnames. Notable exceptions are: (1) both legacy and rich surnames showing strong negative associations with recommendations for entry-level hiring in Thailand, and (2) rich surnames showing strong negative associations with recommendations for leadership positions and loan approvals in Thailand after controlling for perceived power, wealth, intelligence, and commonality.

To ensure robustness, we address the issue of multiple comparisons—which can elevate the risk of false positives (23, 24)—by applying multiple testing corrections to all analyses. The coefficients and adjusted p-values, presented in Tables S1 and S2, confirm that all findings statistically significant at the 1% level remain statistically significant even after these corrections, further supporting the reliability of our results.

To assess the extent to which AI perceptions of surnames mediate the influence of the surnames themselves, we perform a Sobel-Goodman mediation test using the structural equation modeling (SEM) technique. Only mediation results that are statistically significant at the 5% level are included in Table 1.

In both countries, the most prominent pathways associated with legacy or affluent surnames are mediated through the perceived intelligence of individuals bearing these surnames. For example, the effect of having a rich surname on AI recommendations for an executive position is fully mediated by perceived intelligence. Conversely, there is some evidence of a negative mediating effect through perceived wealth in contexts such as entry-level hiring, international school attendance, and political career aspirations. This likely reflects the reality that, in the U.S., individuals from wealthy backgrounds are less likely to pursue entry-level jobs, attend international schools, or enter political careers, leading AI models to adjust their recommendations accordingly.

**Objective Qualifications Mitigate but Do Not Eliminate Surname Bias**

Given that the most significant mediating pathway for the influence of legacy and rich surnames on AI recommendations operates through perceived intelligence, an important question arises: how would the inclusion of additional information—such as the surname holder's qualifications, skill sets, and academic achievements—affect this dynamic? Specifically, would surnames continue to function as a value-added signal beyond these more objective indicators, or would their influence diminish once AI systems are presented with concrete measures of competence?
If surnames retain their predictive power even after controlling for qualifications, it would suggest that AI models implicitly assign intrinsic value to elite surnames beyond their role as proxies for unobserved ability and intelligence. Alternatively, if surname effects weaken in the presence of detailed personal attributes, it would indicate that their influence primarily stems from a lack of available information rather than an inherent bias in AI systems. Understanding this interaction is critical for refining AI-driven decision-making processes, ensuring that recommendations reflect genuine merit rather than inherited status signals.

Figure 5 illustrates coefficient plots of surname categories on perceived intelligence, focusing on the most significant mediating pathway. The analysis includes four conditions: (i) **no profiles**, where only surnames are presented (as shown in Figure 1); (ii) **bad profiles**, in which all surnames are paired with uniformly low qualifications (Honors: No; Tech Skills: No; Special Skills: No; GPA: Not in Top 10%); (iii) **medium profiles**, where all surnames are associated with a mix of qualifications (Honors: No; Tech Skills: Yes; Special Skills: Yes; GPA: Not in Top 10%); and (iv) **good profiles**, where all surnames are paired with consistently high qualifications (Honors: Yes; Tech Skills: Yes; Special Skills: Yes; GPA: in Top 10%).

Across the coefficient plots, we observe that in both countries, the influence of surname categories on AI ratings of intelligence diminishes as more objective information is introduced. For instance, legacy surnames no longer significantly predict perceived intelligence when all surnames in the sample are linked to strong academic profiles, suggesting that AI models rely more on concrete qualifications when available.

There is some suggestive evidence that legacy surnames continue to marginally predict perceived intelligence in Thailand, even when all surnames in the sample are linked to poor academic profiles. For instance, a legacy surname is associated with a 0.087-point increase in perceived intelligence ($p = 0.054$, 95% CI [-0.001, 0.175]). This suggests that, at least in Thailand, when objective qualifications are uniformly low, legacy and wealthy surnames may still function as heuristic indicators of unobserved intelligence, independent of actual skills or academic achievements. In other words, in the absence of other distinguishing credentials, AI appears to default to surname-based assumptions, reinforcing the role of elite surnames as perceived status markers.

# Discussion

"What's in a name?"—a timeless question posed by Shakespeare in *Romeo and Juliet*—takes on new significance in the age of artificial intelligence, where surnames may do more than identify individuals; they may actively shape AI-driven judgments and decisions. This study explores how inherited markers of social status influence algorithmic evaluations across diverse real-world outcomes, even in systems designed to be neutral.

By analyzing real surnames from diverse socioeconomic and historical backgrounds in Thailand and the U.S., we demonstrate that legacy and rich surnames strongly influence AI assessments of power, wealth, and intelligence. These perceptions, particularly those related to intelligence, act as key mediating pathways through which surnames shape AI-driven recommendations in critical areas such as executive hiring, leadership selection, and loan approvals. These findings are consistent with the job market signaling model in economics, which suggests that when direct information about candidates is unavailable, decision-makers rely on observable signals—such as qualifications or, in this case, surnames—as proxies for unobserved abilities and intelligence (15).

As shown in Figure 1, there is strong evidence that legacy and rich variants—surnames phonetically similar to elite surnames but differing in spelling—significantly influence AI perceptions of wealth and intelligence. This suggests that AI, in some cases, treats these variants similarly to genuine elite surnames. However, their impact on AI-driven real-world decisions appears much weaker, indicating that variant surnames do not carry the same weight as genuine legacy and rich surnames.

We also demonstrated that incorporating comprehensive, merit-based information–such as GPA and skill sets–can significantly reduce AI's reliance on surnames as proxies for characteristics such as intelligence, thereby mitigating biases and promoting more equitable outcomes in high-stakes domains such as hiring, leadership selection, and financial decision-making.

It is worth highlighting that Thai surnames exert a more pronounced influence on AI perceptions and judgments compared to U.S. surnames, as reflected in the stronger associations observed in the data. For example, legacy surnames in Thailand are significantly associated with a 0.660-point increase in perceived intelligence, alongside similarly substantial effects on perceptions of wealth and practical judgments, such as leadership appointments and political careers, particularly in contexts where objective qualifications are minimal.

In contrast, while rich and legacy surnames in the U.S. also shape AI perceptions—such as a 0.632-point increase in perceived intelligence for rich surnames—their impact on practical judgments, including hiring decisions and loan approvals, appears less pronounced. This difference may be attributed to Thailand's unique surname law, which mandates that each family has a distinct surname, reinforcing strong associations with status, lineage, and societal standing. In the U.S., by contrast, the widespread sharing of surnames among unrelated families weakens these associations, potentially reducing their influence on algorithmic decision-making.

Our findings have significant societal and ethical implications. First, the ability of AI to assign value and make decisions based on surnames introduces a risk of perpetuating structural inequality within institutions, particularly in contexts involving leadership positions. This suggests the presence of an implicit algorithmic "ceiling" for individuals with more common surnames, potentially restricting their opportunities for advancement or recognition in critical organizational roles from the start.

Second, while the potential for individuals to strategically modify their surnames to closely resemble those of wealthy and influential families might be seen by some to "level the playing field," it raises significant ethical concerns related to authenticity and fairness, particularly when such individuals lack the merit required for the roles they are pursuing.

Third, our findings highlight how AI systems' reliance on surnames as proxies for perceived intelligence and socioeconomic status could further entrench intergenerational inequality stemming from the "birth lottery"—the arbitrary advantage or disadvantage conferred by one's family lineage (25). Additionally, it may reinforce a narrative in which inherited status is perceived as a legitimate indicator of individual merit, fostering the belief that intergenerational inequality is both natural and acceptable (26, 27). This has profound implications for social mobility, as individuals without elite lineage may face systemic barriers to advancement, even in contexts explicitly designed to be meritocratic (31, 32). By embedding historical advantages into decision-making processes, AI risks institutionalizing disparities that erode equal opportunity and limit upward mobility.

Fourth, it is possible that surname-based biases that influence hiring recommendations or leadership selection can cascade into other aspects of institutional operations, such as mentorship opportunities, salary decisions, or performance evaluations. Real-world solutions must be pursued urgently to counteract these effects. Our findings issue an urgent call for developers, policymakers, and institutions to take ethical responsibility in scrutinizing the sociocultural assumptions embedded within AI systems (31-36). This responsibility goes beyond addressing algorithmic biases, encompassing a comprehensive evaluation of the objectives, metrics, and data that underpin AI performance. To ensure that technological progress advances broader societal goals of fairness and inclusivity, AI systems must be intentionally designed to challenge and dismantle structural inequalities rather than inadvertently perpetuate them.

Our findings on surname bias in AI decision-making also have significant practical implications beyond academic significance. For instance, despite existing laws and policies aimed at promoting equity for underrepresented groups—such as affirmative action and diversity, equity, and inclusion (DEI) initiatives—there are no explicit legal safeguards against discrimination based on family lineage as inferred from surnames. If AI-driven or human decision-making systematically exclude individuals due to their surnames, this regulatory gap may warrant new policies that address such biases. Recognising surname bias could help policymakers refine anti-discrimination laws, such as the European Union's AI Act, and develop clearer guidelines for AI accountability to ensure fairness in automated decision-making systems. Additionally, addressing surname bias is also crucial for organisations that integrate AI into their recruitment and decision-making processes. AI-driven decisions in hiring, lending, and public services could inadvertently favor or disadvantage individuals based on surname associations, potentially exposing companies to legal challenges. If candidates or customers can demonstrate that they were unfairly excluded due to algorithmic bias, organisations may face lawsuits citing unfair discrimination. Proactively mitigating surname bias in AI systems is therefore not only an ethical responsibility but also a strategic necessity for reducing legal and reputational risks.

Why might AI preferentially favor elite surnames in its decision-making process? Several factors could contribute to this bias. First, historical training data may embed long-standing social biases, associating certain surnames with higher success rates and prompting algorithms to reinforce these patterns (28, 29). Second, the socioeconomic advantages linked to elite surnames—such as access to prestigious education or influential networks—may further skew AI-driven decisions. Third, AI models may engage in automated stereotyping, detecting superficial correlations rather than genuine indicators of merit, thereby codifying existing inequities. Finally, the opacity of many AI systems could make it challenging to detect and correct these biases, allowing them to persist over time. Future research should investigate these potential underlying mechanisms behind AI's preferential treatment of elite surnames to develop more equitable decision-making frameworks.

Despite our work's contributions, the study is not without potential objections. For example, the AI systems evaluated, such as ChatGPT, relied on specific training data and algorithms, which may not fully capture the variability in models and methods currently in use. However, other studies have provided evidence that existing AI systems—e.g., GPT4, Claude 2, Gemini Pro, and GPT-3.5—tend to exhibit the same human biases (30). In addition, while the choice of real-world outcomes for which AI had to make judgments—such as hiring recommendations, loan approvals, and leadership appointments—may appear arbitrary, they were chosen deliberately due to their societal significance and their alignment with domains where AI-driven decisions are increasingly prevalent. Nevertheless, we acknowledge that other domains with a history of systemic inequality, such as university admissions and access to affordable housing, could have been considered as well.

Another potential objection to our findings is that AI's reliance on rich and legacy surnames as proxies for intelligence stems from the empirical association between wealth and human capital accumulation—where individuals from affluent backgrounds are more likely to have had greater access to education and skill development, making them statistically more intelligent than others. From this perspective, the observed bias may not simply be a form of taste-based discrimination but rather an instance of statistical discrimination.

While this argument may hold in some cases, our evidence suggests that affluent and prestigious surnames exert an independent influence on AI's assessment of intelligence, even when all candidates exhibit uniformly poor academic performance. This finding implies that AI does not merely associate surnames with access to superior education but rather uses them as proxies for intrinsic intelligence—an assumption that is not universally valid.

Furthermore, distinguishing between taste-based and statistical discrimination in AI evaluations is inherently challenging. However, relying on surnames as a heuristic for intelligence can create a self-reinforcing cycle: individuals with common surnames may be discouraged from investing in education or skill development, knowing that their expected returns in the job market are lower than those with elite surnames. Over time, this dynamic could exacerbate existing social inequalities, underscoring the need for targeted interventions to mitigate surname-based biases in AI-driven decision-making.

In conclusion, this study highlights the risks of algorithmic bias in reinforcing social stratification and limiting social mobility. In meritocratic societies, the automation of intergenerational privilege not only restricts opportunities for individuals with common surnames but also reinforces a narrow and flawed view of merit. The consequences are profound: exclusion from leadership roles, career advancement, and financial opportunities based on an immutable marker of identity. Such biases risk deepening existing inequalities and eroding public trust in both AI technologies and the institutions that deploy them.

The implications extend beyond the technical domain, demanding a broader societal reckoning with the intersection of technology, privilege, and discrimination. Addressing these challenges is not merely a technical necessity but a moral imperative. Without intervention, AI systems may entrench a world where inherited advantages are mistaken for personal merit. The path forward requires greater transparency, robust regulation, and a fundamental rethinking of how merit is defined and rewarded in the digital age.

**Limitations & Future Works**

Our study has several limitations that future research should address. First, by focusing solely on surnames from the U.S. and Thailand, our findings may not generalize to cultures with different surname conventions. Second, while we identify surname-based biases, we do not examine their interactions with other identity markers such as first names, gender, or race, highlighting the need for larger, more diverse datasets and simulation studies. Third, our analyses rely on simulated scenarios rather than real-world contexts, underscoring the importance of controlled experiments and longitudinal studies to assess how human decision-makers respond to AI recommendations. Finally, while our mediation analysis identifies perceived intelligence, power, and wealth as key mechanisms, other factors—such as trustworthiness or cultural familiarity—may also play a role and warrant further investigation. Future research should expand cultural and identity dimensions, incorporate field experiments, and explore a broader range of mediators to develop more transparent, equitable, and robust AI systems.

**Materials and Methods**

Our study investigates AI bias in surname-based decision-making using a robust methodology encompassing dataset creation, name variation generation, and systematic evaluation. We curated a dataset comprising 300 surnames across three categories—rich, legacy, and common—selected from Thailand and the United States, with 50 surnames per category per country. These surnames were sourced from Forbes lists, historical archives, and population databases. To account for linguistic diversity, phonetically plausible variants were generated for each surname using language-specific character substitutions, resulting in a total of 600 surnames.

The experimental design utilized four configurations: (1) surnames presented in isolation and (2) surnames paired with controlled profiles representing three qualification levels (good, medium, and bad). Evaluations were performed using GPT-4o-mini, the leading language model in 2024. Each surname was analyzed across ten socioeconomic dimensions, with three independent evaluations conducted for every surname-dimension pairing in each experimental condition. This comprehensive approach yielded 72,000 evaluations—600 surnames (including variants) × 10 dimensions × 3 evaluations per dimension × 4 candidate profile configurations (good, medium, bad, and retracted). Data collection was performed via a standardized API-based protocol to ensure rigor and reproducibility.

**Surname Datasets**

To investigate potential AI bias related to surnames, we developed a dataset encompassing three distinct categories of last names from Thailand and the United States: rich surnames, legacy surnames, and common surnames. Each category included 50 surnames per country, resulting in a total of 300 surnames. Rich surnames were randomly sampled from the 2024 and 2025 Forbes lists of the richest individuals in each country. For Thailand, legacy surnames were selected from royally bestowed names and those associated with relatives of the royal family, reflecting the historical prominence of royal-affiliated families in governance and business leadership prior to Thailand's democratic transition in 1923. In the U.S., legacy surnames were drawn from well-known dynastic families in politics, entertainment, and business. Common surnames were randomly selected from the most frequently occurring surnames in each country's population databases.

The Thai surname categories were completely distinct, with no surnames appearing in multiple categories. In contrast, the U.S. dataset exhibited some overlap between categories. Specifically, four surnames—Walton, Scott, Johnson, and Roberts—appeared in both the rich and common categories, while another four—Jackson, Adams, Lee, and Carter—were present in both the legacy and common categories. This overlap highlights the dynamic nature of social mobility and wealth accumulation in the United States, where some historically common surnames have become associated with significant wealth or dynastic influence over time. The complete list of surnames used in this study is provided in Appendix A.

**Surname Variations**

Building on the original dataset, we generated a set of plausible surname variations through systematic character substitutions (detailed in Appendix C) designed to preserve phonetic similarity and adhere to orthographic conventions. For each surname, variations were created by randomly substituting a single character at a random position, following predefined substitution rules specific to the linguistic characteristics of each language.

For Thai surnames, we applied substitutions using visually or phonetically similar Thai characters. These included consonants with comparable articulation points (e.g., ค/ก, น/ณ, ธ/ท, พ/ภ, ส/ศ), vowel length variants (e.g., ะ/า, ิ/ี), and common phonological alternations (e.g., ร/ล, ซ/ส). The substitutions were carefully designed to adhere to Thai phonotactic constraints and orthographic conventions, ensuring that the generated variants remained plausible as Thai surnames.

For English surnames, we implemented two levels of substitutions: digraph-level and single-character replacements. Digraph substitutions included common spelling variants found in English surnames (e.g., 'ph'/'f,' 'th'/'t,' 'll'/'l'), while single-character replacements focused on frequent orthographic variations (e.g., 'c'/'k,' 'f'/'ph,' 's'/'z,' 'v'/'w') and vowel pattern changes (e.g., 'ee'/'ea,'

'ie'/'y,' 'yn'/'in'). These substitution patterns were based on documented variations in English surname orthography, ensuring the generated variants adhered to plausible phonological and spelling conventions. A complete list of surname variants used in the study is provided in Appendix B.

**Experimental Datasets**

The complete dataset comprises 300 original surnames (50 surnames × 3 categories × 2 countries) and their 300 phonetically plausible variants, resulting in 600 unique entries. These surnames were used to construct four distinct experimental datasets. Following established methodologies in candidate selection bias research (2), each profile included four standardized fields: GPA, Honors, Technical Skills, and Special Skills. The four experimental configurations were designed to isolate the effect of surnames while systematically controlling for qualification variables:

The experimental datasets were structured as follows:

1. **"Only Surname" Dataset**: This baseline dataset includes only the surnames, with no additional profile information. It enables a direct examination of surname-based bias independent of any qualifications.

2. **"Good Profile" Dataset**: Each surname is paired with uniformly high qualifications, including Honors (Yes), Tech Skills (Yes), Special Skills (Yes), and GPA (Top 10%). This configuration assesses whether surname-based bias persists even when profiles demonstrate consistently strong credentials.

3. **"Medium Profile" Dataset**: Each surname is combined with mixed qualifications: Honors (No), Tech Skills (Yes), Special Skills (Yes), and GPA (Not in the Top 10%). This intermediate setup investigates how surname-based bias interacts with a more nuanced qualification profile.

4. **"Bad Profile" Dataset**: Each surname is associated with uniformly low qualifications, including Honors (No), Tech Skills (No), Special Skills (No), and GPA (Not in the Top 10%).

This configuration explores whether surname-based bias exacerbates or diminishes the impact of weaker credentials.

**Candidate Evaluation Experiments**

To assess potential biases associated with surnames, we conducted systematic evaluations using GPT-4o-mini, the industry-leading language model in 2024. Renowned for its superior performance and computational efficiency, GPT-4o-mini has been widely adopted by major technology companies and research institutions, making it an ideal tool for examining real-world AI biases that could impact automated screening systems.

Each surname was evaluated across ten dimensions relevant to socioeconomic perception and decision-making: perceived power, perceived wealth, perceived intelligence, leadership potential, executive-level hiring, entry-level hiring, international school admission, political career prospects, loan approval likelihood, and name commonality. Scores were assigned on a scale from 0 to 10. Detailed evaluation prompts are provided in Appendix D1 and D2. To ensure unbiased evaluations, the model assessed each surname independently, without memory or context of previous assessments.

To enhance robustness and account for potential variability in model outputs, three independent evaluations were conducted for each surname-dimension combination, repeated across four rounds for the four experimental datasets: "Only Surname," "Good Profile," "Medium Profile," "Bad Profile." This setup enabled the calculation of mean scores while controlling for stochastic variations in the model's responses. The evaluation prompts were designed to replicate real-world decision-making scenarios, such as:

- "As a hiring manager, rate how likely you would shortlist this candidate for an executive position in Thailand. (0 = least likely, …, 10 = most likely.)"
- "As a loan officer, rate how likely you would approve this person's loan application in Thailand. (0 = least likely, …, 10 = most likely.)"

The prompts were consistently structured across all evaluations, with only the surname (and associated profile, where applicable) varying. This controlled design allowed us to isolate the effect of surnames on the model's perception and decision-making processes, minimizing the influence of extraneous variables.

Evaluations were conducted via API calls to GPT-4o-mini, using a JSON-structured output format to ensure consistent, quantifiable responses across all assessments. The full API call structure and response schema are detailed in Appendix E. By eliminating comparative context and ensuring the model had no memory of prior assessments, we ensured that each evaluation decision was made in isolation, providing a clear measure of surname-based bias.

**Statistical Analysis**

We estimated ordinary least squares (OLS) on the following regressions:

$$P_i = \beta SN_i + u_i \quad (1)$$
$$D_i = \gamma SN_i + \epsilon_i, \quad (2)$$
$$D_i = \delta P_i + \gamma SN_i + \epsilon_i, \quad (3)$$

where $P_i$ represents AI perceptions of surname *i*, including power, wealth, intelligence, and commonality; $D_i$ is AI recommendations for different real-world outcomes, including executive hire, leadership, entry hire, international school, political career, and loan approval; $SN_i$ represents surname categories, including legacy surnames, legacy variants, rich surnames, rich variants, and common variants. The $u_i$ and $\epsilon_i$ are the error terms. Bootstrap standard errors were computed with 1,000 replications and reported for all estimations. Figure 1 displays the coefficients estimated from Equation 1, while those from Equation 2 are presented in Figure 2. Figures 3 and 4 illustrate the coefficients derived from Equation 3.

For the Sobel-Goodman test, we employed structural equation modeling (SEM) based on Equation 3 and used STATA's *medsem* command to calculate the indirect effects and their corresponding p-values. For the adjustment of p-values due to multiple hypothesis testing, we applied STATA's *wyoung* command to the equations.

## Acknowledgments

**Funding:** MIT Media LabConsortium Lab Member (CLM)

**Author contributions:**
- Conceptualization: NP, PP
- Methodology: NP,PP
- Investigation: NP
- Visualization: PP
- Supervision: PM
- Writing—original draft: NP,PP
- Writing—review & editing: NP,PP,PM

**Competing interests:** Authors declare that they have no competing interests

**Data and materials availability:** All data, code, and materials used in the analyses is available at https://github.com/mitmedialab/powerful_names

# Figures and Tables

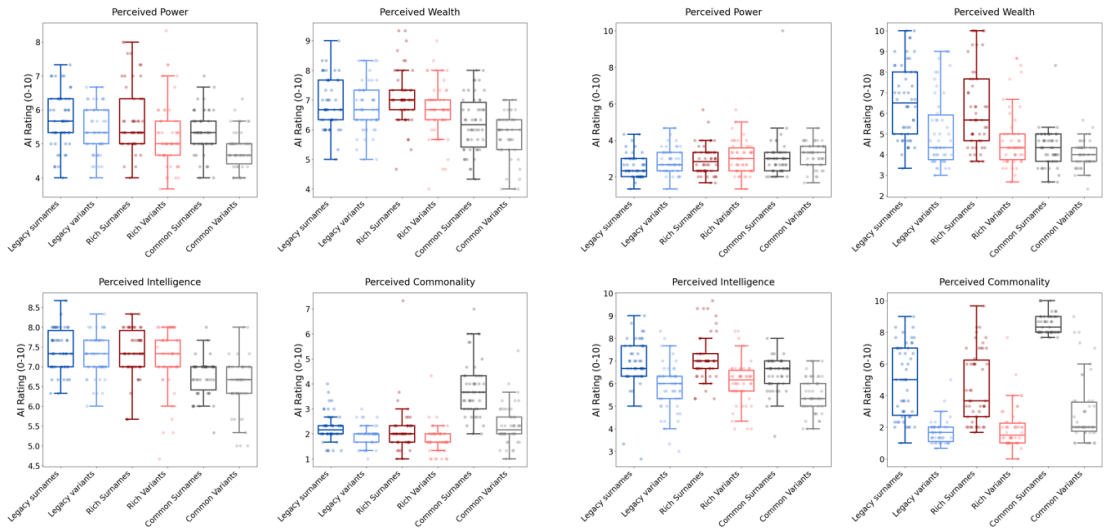

A: Box plot--Thailand          B: Box plot--the U.S.

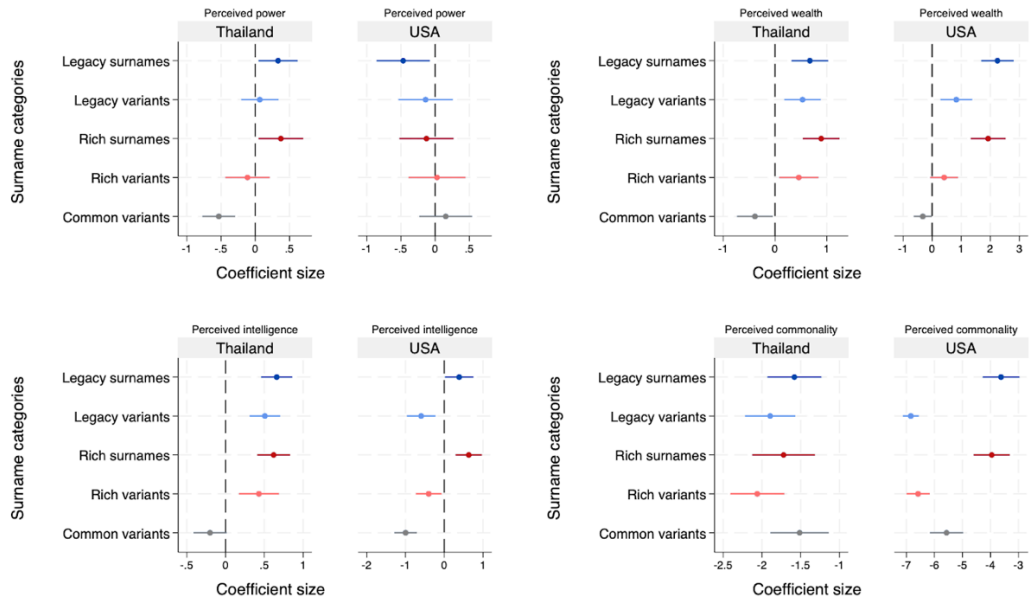

C. Coefficient plots--Thailand and the U.S.

**Fig. 1**. Box plots and coefficient plots illustrating the relationships between surname categories and AI-generated ratings for power, wealth, intelligence, and commonality. *Panel A* depicts the distribution of AI ratings across surname categories for Thailand. *Panel B* presents the corresponding distributions for the U.S. *Panel C* reports

coefficient plots for both countries, highlighting the effect sizes of surname categories on the evaluated dimensions. The dataset includes 600 surnames (N = 300 each from Thailand and the U.S.), with ratings measured on a 0–10 scale (0 = lowest, 10 = highest). The 95% confidence intervals are calculated using bootstrap standard errors with 1,000 replications. Common surnames serve as the reference category in the analysis.

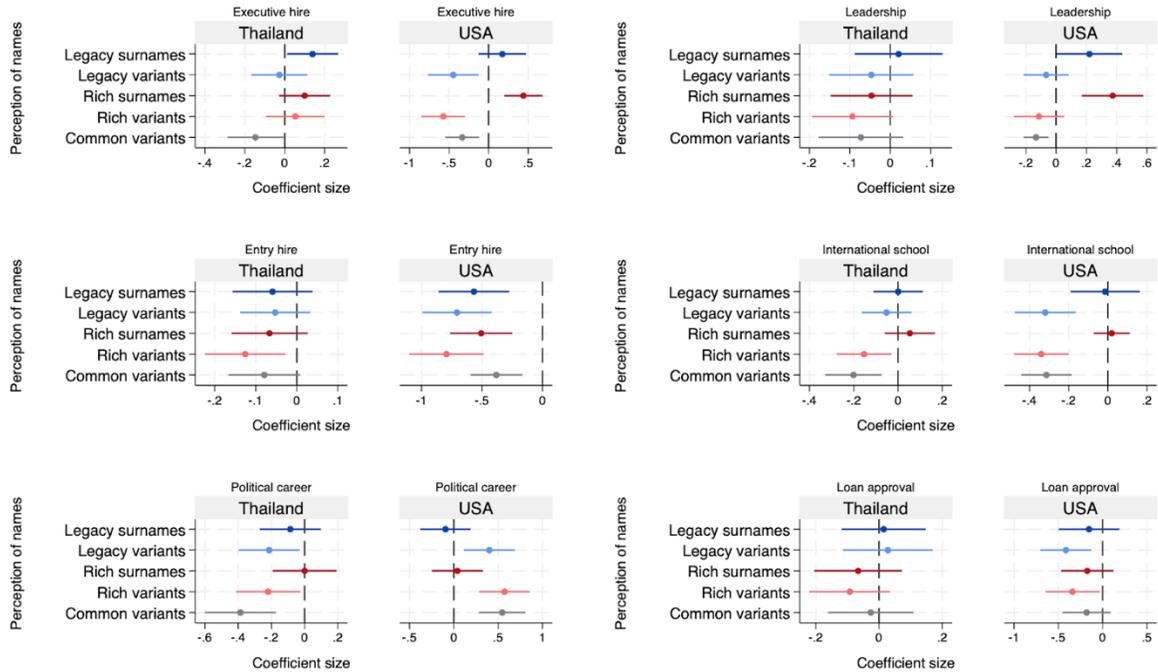

**Fig. 2.** Coefficient plots illustrating the relationships between surname categories and AI judgment in various real-world scenarios. These scenarios include recommendations for executive hiring, entry-level hiring, political candidacy, leadership roles, international school admissions, and loan approvals. Scenario responses are measured on a scale from 0 (strongly not recommended) to 10 (strongly recommended). The 95% confidence intervals are calculated using bootstrap standard errors with 1,000 replications.

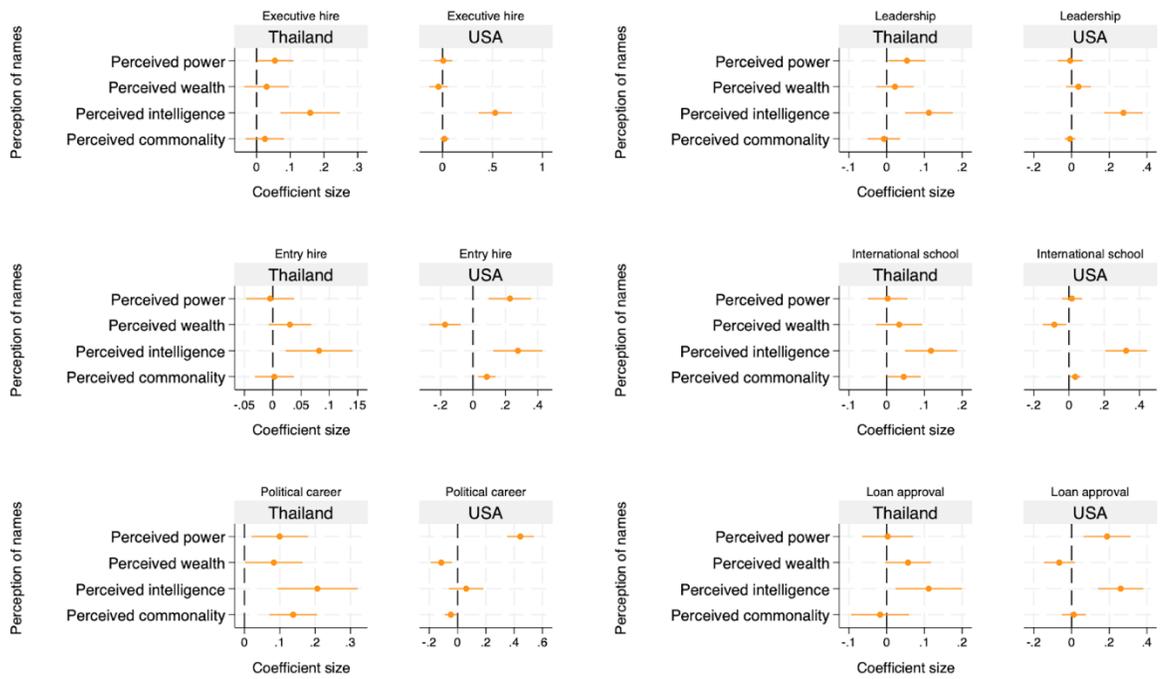

**Fig. 3.** Coefficient plots showing the partial correlations between AI-generated ratings (power, wealth, intelligence, and commonality) and AI judgment in various real-world scenarios. These scenarios include recommendations for executive hiring, entry-level hiring, political candidacy, leadership roles, international school admissions, and loan approvals. Scenario responses are measured on a scale from 0 (strongly not recommended) to 10 (strongly recommended). All regression models include dummies representing surname categories as control variables. The 95% confidence intervals are calculated using bootstrap standard errors with 1,000 replications.

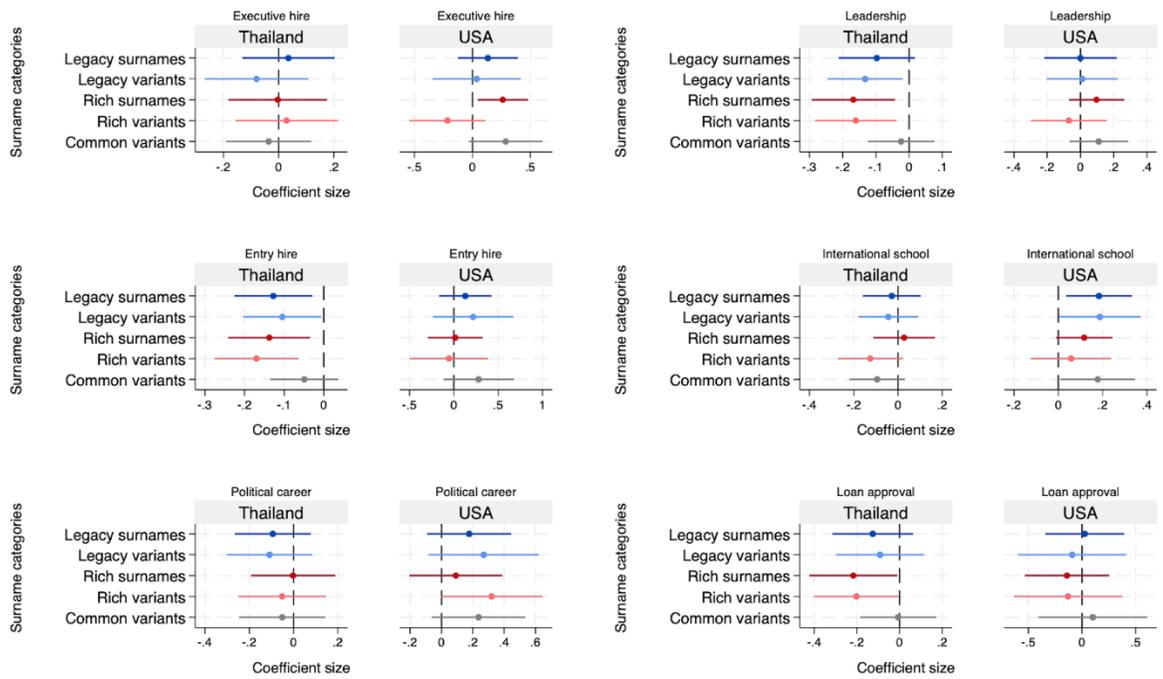

**Fig. 4.** Coefficient plots showing the partial correlations between surname categories and AI judgment in various real-world scenarios. All regression models include perceived power, wealth, intelligence, and commonality as control variables. The 95% confidence intervals are calculated using bootstrap standard errors with 1,000 replications.

**Table 1:** Sobel-Goodman Test Results for Legacy and Rich Surname Influences within the SEM Framework

|  | Coeff. | Sobel p-value |
|---|---|---|
| **Thailand** | | |
| **1) Executive hire** | | |
| Legacy -> Perceived intelligence | 0.105 | 0.002 |
| Rich-> Perceived intelligence | 0.099 | 0.002 |
| **2) Leadership** | | |
| Legacy-Perceived intelligence | 0.074 | 0.005 |
| Rich-> Perceived intelligence | 0.069 | 0.005 |
| **3) Entry hire** | | |
| Legacy -> Perceived intelligence | 0.054 | 0.015 |
| Rich-> Perceived intelligence | 0.051 | 0.016 |
| **4) International school** | | |
| Legacy -> Perceived intelligence | 0.078 | 0.004 |
| Rich-> Perceived intelligence | 0.073 | 0.004 |
| **5) Political career** | | |
| Legacy -> Perceived intelligence | 0.136 | 0.002 |
| Rich-> Perceived intelligence | 0.128 | 0.002 |
| **6) Loan approval** | | |
| Legacy -> Perceived intelligence | 0.074 | 0.016 |
| Rich-> Perceived intelligence | 0.069 | 0.018 |
| **U.S.** | | |
| **1) Executive hire** | | |
| Rich-> Perceived intelligence | 0.265 | 0.022 |
| **2) Leadership** | | |
| Rich-> Perceived intelligence | 0.139 | 0.026 |
| **3) Entry hire** | | |
| Legacy -> Perceived wealth | -0.271 | 0.005 |
| Rich->Perceived wealth | -0.205 | 0.014 |
| **4) International school** | | |
| Legacy->Perceived wealth | -0.131 | 0.035 |
| Rich-> Perceived intelligence | 0.163 | 0.021 |
| **5) Political career** | | |
| Legacy->Perceived wealth | -0.181 | 0.013 |
| Rich->Perceived wealth | -0.137 | 0.025 |
| **6) Loan approval** | | |
| Rich->Perceived intelligence | 0.131 | 0.034 |

**Note**: p-values are calculated using bootstrap standard errors with 1,000 replications.

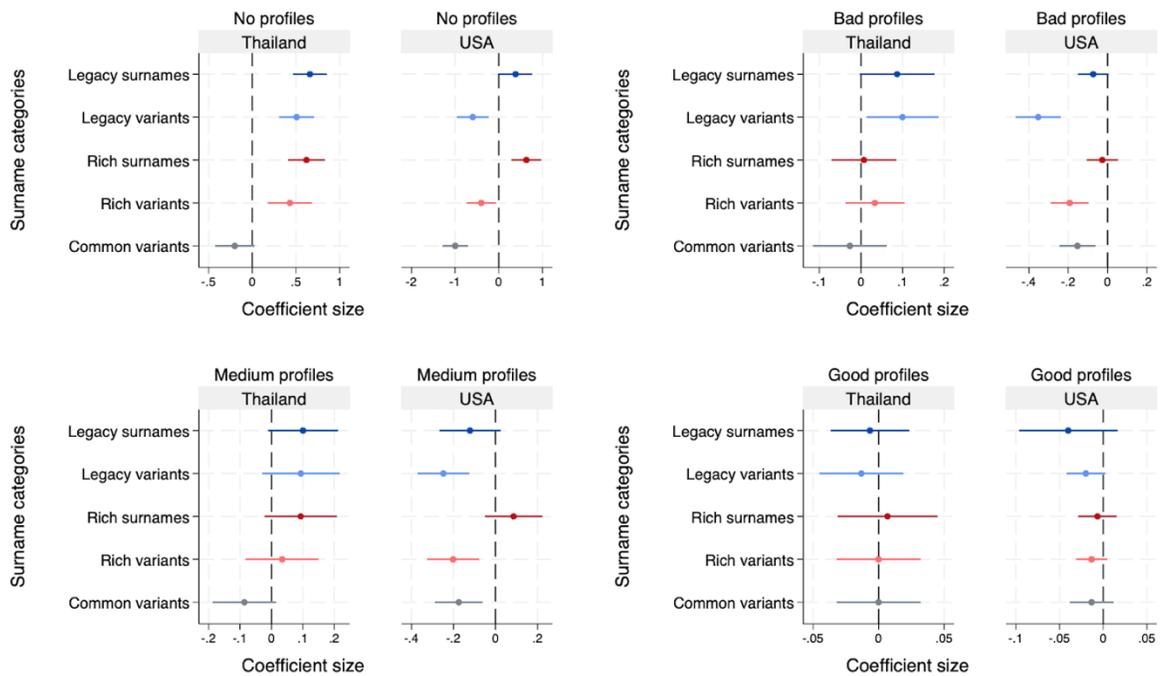

**Fig. 5.** Coefficient plots showing the association between surname categories and AI-generated rating of intelligence across profile samples. The 95% confidence intervals are derived from bootstrap standard errors with 1,000 replications. Common surnames are used as the reference category.